\documentclass[12pt]{article}

\usepackage[a4paper, margin=1in]{geometry}
\usepackage{graphicx}   
\usepackage{amsmath}    
\usepackage{amssymb}    
\usepackage{setspace}   
\usepackage{titlesec}   
\usepackage{natbib}     
\usepackage{hyperref}   
\usepackage{multirow}
\usepackage{algorithm}
\usepackage{algorithmicx}
\usepackage{algpseudocode}
\usepackage{pdflscape}
\usepackage{booktabs,adjustbox,pdflscape}
\usepackage{array}
\usepackage{times}
\usepackage{caption}   
\usepackage{float}     

\titleformat{\section}{\normalfont\large\bfseries}{\thesection}{1em}{}
\titleformat{\subsection}{\normalfont\normalsize\bfseries}{\thesubsection}{1em}{}
\titleformat{\subsubsection}[runin]{\normalfont\normalsize\bfseries}{\thesubsubsection}{1em}{}

\title{Optimal Portfolio Construction - A Reinforcement Learning Embedded Bayesian Hierarchical Risk Parity (RL-BHRP) Approach}
\author{
    Shaofeng Kang \thanks{University of Toronto} \\
    \texttt{shaofeng.kang@mail.utoronto.ca}
    \and
    Zeying Tian \thanks{McGill University} \\
    \texttt{zeying.tian@mail.mcgill.ca}
}
\date{\today}

\begin{document}

\maketitle
\thispagestyle{empty}

\vspace{2em}

\begin{abstract}
    We propose a two-level, learning-based portfolio method (RL-BHRP) that spreads risk across sectors and stocks, and adjusts exposures as market conditions change. Using U.S. equities from 2012 to mid-2025, we design the model using 2012–2019 data and evaluate it out-of-sample from 2020 to 2025 against a sector index built from exchange-traded funds and a static risk-balanced portfolio. Over the test window, the adaptive portfolio compounds wealth by approximately 120\%, compared with 101\% for the static comparator and 91\% for the sector benchmark. The average annual growth is roughly 15\%, compared to 13\% and 12\%, respectively. Gains are achieved without significant deviations from the benchmark and with peak-to-trough losses comparable to those of the alternatives, indicating that the method adds value while remaining diversified and investable. Weight charts show gradual shifts rather than abrupt swings, reflecting disciplined rebalancing and the cost‑aware design. Overall, the results support risk-balanced, adaptive allocation as a practical approach to achieving stronger and more stable long-term performance.

\end{abstract}

\vspace{1em}

Portfolio Optimization, Asset Allocation 

\newpage
\section{Introduction}
Modern Portfolio Theory (MPT) provides a mathematical framework for allocating assets to maximize returns while maintaining a certain level of risk across asset classes. Since Markowitz proposed this mean-variance analysis, institutional and retail investors have sought systematic ways to allocate capital across hundreds of securities. Although MPT gained popularity over the past decades, its ineffectiveness of estimating expected returns have always been problematic to both industry and academia. Traditional optimization relies on point estimates of expected returns and covariances, yet these inputs are notoriously noisy.  Estimation error propagates into extreme, highly leaped weights and poor out-of-sample performance, an effect sometimes called 'error maximization'. 
\subsection{Risk Based Portfolios}
To solve the problem of a small forecasting error in expected returns leading to a significant changes in asset allocations, \cite{maillard2008properties} introduced equal weighted risk portfolio to overcome the ineffectiveness in MPT, where they assign weights computed by solving the covariance-matrix–based system that equalizes each asset’s marginal contribution to total portfolio variance to achieve optimality. \cite{lopez2016building} introduced the Hierarchical Risk Parity (HRP) framework to reduce errors from inverting ill-conditioned covariance matrices, and to assign capital by recursively equal-risk splitting at each node rather than at the individual-asset level. Consequently, HRP produces more diversified and stable allocations that have repeatedly outperformed standard benchmarks in empirical and simulated studies, offering practitioners a practical and scalable alternative to classical optimization methods. 
\subsection{Bayesian Enhanced Portfolios}
Bayesian-based methods have been studied and implemented in various academic research. \cite{cooper2020black} replaces the traditional capitalization prior in Black-Litterman optimization with a Risk Parity (RP) portfolio that equalizes the contribution to total variance. They argue that RP could be more efficient than the capital‑weighted market portfolio and is feasible when market capitalization weights do not exist (e.g., hedge funds, commodities).
\subsection{Deep Reinforcement Learning (DRL)}
\cite{millea2022using} deploys a two-tier architecture in which a Proximal Policy Optimization (PPO) agent supervises twelve Hierarchical Risk Parity (HRP) and Hierarchical Equal Risk Contribution (HERC) allocators plus a cash action. Each allocator, distinguished by its covariance look-back window, runs offline; their rolling one-hour returns form the state vector for the PPO, converting continuous weight selection into a discrete model-choice problem. This structure reframes portfolio construction as a nonstationary multi-armed bandit, sidestepping the exploration burden of high-dimensional action spaces. The reward couples absolute portfolio growth with an incentive forchoosing cash when all assets decline, codifying disciplined risk-off behavior. After a single training pass that begins 700 hours into each market series, the learned policy remains fixed over an equally long out-of-sample window, eliminating retraining costs. Across cryptocurrency, equity, and FOREX datasets, the hierarchical agent consistently outperforms every constituent allocator, a random selection baseline, and equal-weight buy-and-hold portfolios in terminal wealth and volatility.
\section{Motivation}
Modern portfolio construction techniques face persistent trade-offs for both institutional and retail investors. Professional practitioners and academic researchers aim to develop a method that can amplify the benefits of each approach while mitigating the drawbacks of trade-offs. Traditional mean-variance approaches lack stability in forecasting expected returns. Risk-parity frameworks deliver stable, well-diversified allocations, but treat expected returns as noise, leaving performance on the table when market premia shift. With the recent advancement of DRL, reinforcement learning-based DRL allocators can react to regime changes; yet, without explicit risk budgets, they often generate concentrated, high-turnover positions that breach institutional limits. Existing attempts to merge the two paradigms address only part of the problem. Firstly, static Bayesian-risk-parity models, such as Bayesian Risk Parity (BRP), update the mean estimate once and then revert to a standard optimization loop; they do not learn online or enforce hierarchy across sectors. Secondly, DRL “switchers” supervise several pre‑computed HRP/HERC portfolios but can only jump from one static weight vector to another; the RL layer never modifies risk budgets or incorporates priors. Thirdly, transaction costs, or the cost of rebalancing, are usually omitted or underestimated. This underestimation could lead to frequent rebalances, making the strategy practically impossible. In our research, we proposed the RL-BHRP framework, which unifies Bayesian estimation, risk-parity risk control, and policy-gradient learning in a single loop.

\section{Methodology}
We consider a dynamic portfolio optimization problem in the equity market, where an agent sequentially allocates capital among tradable assets defined the investor to maximize long-term risk-adjusted returns.  To incorporate informed views on returns and risks, we impose a Bayesian hierarchical prior structure over asset returns and volatilities at the sector level. Specifically, assets are grouped into sectors $g=1,...,G$ (e.g. industry sectors). We assume each sector has latent parameters governing its expected return and risk, which in turn inform the priors for individual stocks in that sector. For example, let $\mu_{g}$ be the prior mean return for sector $g$ and $\mu_{i}$ for stock $i$ in sector $g$. A hierarchical model may assume $\mu_{g} \sim \mathcal{N}(\mu_{\text{market}}, \sigma^2_{\mu,g})$ and $\mu_{i} \sim \mathcal{N}(\mu_{g}, \sigma^2_{\mu,i})$, so that stock-level expectations are shrunk toward a sector-level mean. Likewise, covariance estimations can be structured: one may use sector-specific factors or shrinkage of individual asset volatilities toward a sector volatility level. \cite{feng2022factor} demonstrated that a multi-level Bayesian prior encourages information sharing across assets within a sector, improving estimation robustness, especially for assets with scarce data. \cite{ninsinadvanced} demonstrated that Bayesian hierarchical structures offer a flexible approach to capturing layered uncertainty, time-varying covariances, and shared sectoral effects. 

\subsection{BHRP Model}

Assume that we work on a filtered probability space $(\Omega, \mathcal{F}, \{\mathcal{F}_t\}_{t\geq 0}, \mathbb{P})$. The complete probability space $(\Omega, \mathcal{F}, \mathbb{P})$ is equipped with a continuous right filtration and increases $\{\mathcal{F}_t\}_{t\geq 0}$. With regard to asset returns, we assume there are $N$ assets, $G$ sectors, and $T$ trading days. For each asset $i$ in the index set $\mathcal{I}$, $i\in \{1,\ldots,N\}$, the one period return process $\{R_{i, t+1}\}_{t\geq 0}$ is $\mathcal{F}_{t+1}-$ measureable and adapted, with $$\sup_{t,i} |R_{i, t+1}| \leq L_R < \infty \quad a.s.$$  For each sector $j$ in the index set $\mathcal{G} = \{1,\ldots, G\}$, we have a surjective map from assets to sectors $$g: \mathcal{I}\to \mathcal{G}$$ assigns each asset to a sector. 
\subsubsection{Definition: Tradable Universe} Define a subset of assets $I_\lambda\subseteq \mathcal{I}$ such that $I_\lambda = \{i: ADV_i \geq \overline{adv}, P_i \geq \overline{p}, i\in \mathcal{I}\}$. That is, the 20-day average daily dollar volume exceeds 10 million USD, and the closing price is greater than 3 USD. The $I_\lambda$ set is re-screened quarterly, and the resulting set is fixed for the subsequent quarter to avoid any look-ahead bias.  
\subsubsection{Definition: Hierarchical Composition} Each sector $g$ holds a fixed subset $I_g$ of stocks such that $I_g = \{i\in I_\lambda: g(i) = g\}$. If we select top $K$ rank of $I_g$, that is to truncate $|I_g|$ to the largest $K$ stocks if $|I_g| > K$.\\
\\
The RL-BHRP Model framework consists of three main components. Specifically, the Bayesian Hierarchical Return modeling, the risk parity component for total risk contribution, and the Reinforcement Learning agent that facilitates dynamic asset allocation. 
\subsubsection{Definition: Market Level Mean $\mu_M$} $\mu_M\sim \mathcal{N}(0, \sigma_M^2)$ captures the economy-wide or market-wide drift. The variance expresses the prior uncertainty about the drift. 
\subsubsection{Definition: Sector Level Mean $\mu_g$} For each sector $g\in \mathcal{G}$, $\mu_g | \mu_M \sim \mathcal{N}(\mu_m, \sigma_g^2)$. Theoretically, the sector mean should be derived from the market drift. Larger $\sigma_g^2$ allows sectors to deviate further from the market trend. 
\subsubsection{Definition: Asset Level Mean $\mu_i$} For each asset $i\in I_\lambda$, $\mu_i | \mu_{g(i)}\sim \mathcal{N}(\mu_{g(i)}, \sigma_{g(i)}^2)$. 
\subsubsection{Definition: Conditional Returns} Given the return process $\{R_{i, t+1}\}_{t\geq 0}$, the conditional return given asset level mean distribution is $R_{i, t+1}|\mu_i \sim \mathcal{N}(\mu_i, \sigma_i^2)$ \\
\\
\subsubsection{Weights Model}
Let $\mathcal{I}=\{1,\dots,N\}$ be the asset set and $\mathcal{G}=\{1,\dots,G\}$ the set of sectors.
Each asset $i\in\mathcal{I}$ belongs to exactly one sector via a surjective map $g:\mathcal{I}\to\mathcal{G}$.
For $g\in\mathcal{G}$, let $\mathcal{I}_g=\{i\in\mathcal{I}: g(i)=g\}$ and $n_g=|\mathcal{I}_g|$.
For $K\ge1$ define the (long‑only, fully‑invested) $K$‑simplex
$$
\Delta_K \;=\; \bigl\{\,x\in\mathbb{R}^K_+:\; \mathbf{1}^\top x=1\,\bigr\},
\qquad
\mathrm{int}(\Delta_K)=\{x\in\Delta_K:\; x_k>0\ \forall k\}.$$
\\
The sector‑level weights are a vector $W=(W_1,\ldots,W_G)\in\Delta_G$.
For each sector $g$, the within‑sector weights are $\eta^{(g)}=(\eta_{i\mid g})_{i\in\mathcal{I}_g}\in\Delta_{n_g}$.

Define 
$$\Phi:\ \Delta_G\times\prod_{g=1}^G \Delta_{n_g}\to \Delta_N$$ 
by 
$$w_i \;=\; W_{g(i)}\,\eta_{i\mid g(i)},\qquad i\in\mathcal{I_\lambda}$$

The resulting $w=(w_1,\dots,w_N)$ are the final asset‑level weights.

\subsubsection{Proposition: Feasibility and Interior Bijection}
{\rm (a)} For any $(W,\{\eta^{(g)}\})$ in the domain of $\Phi$, one has $w=\Phi(W,\eta)\in\Delta_N$.
{\rm (b)} The restriction $\Phi:\ \mathrm{int}(\Delta_G)\times\prod_{g}\mathrm{int}(\Delta_{n_g})\to \mathrm{int}(\Delta_N)$ is a bijection with inverse
\[
W_g(w)\;=\;\sum_{i\in\mathcal{I}_g} w_i,
\qquad
\eta_{i\mid g}(w)\;=\;\frac{w_i}{W_g(w)}\quad (i\in\mathcal{I}_g).
\]

Proof:
(a) Non‑negativity is immediate. For the budget constraint,
\[
\sum_{i=1}^N w_i \;=\; \sum_{g=1}^G \sum_{i\in\mathcal{I}_g} W_g\,\eta_{i\mid g}
\;=\; \sum_{g=1}^G W_g\Bigl(\sum_{i\in\mathcal{I}_g}\eta_{i\mid g}\Bigr)
\;=\; \sum_{g=1}^G W_g \;=\; 1.
\]
(b) If $W_g>0$ and $\eta_{i\mid g}>0$ for all $g,i$, then $w_i>0$ and $w\in\mathrm{int}(\Delta_N)$. Conversely, given $w\in\mathrm{int}(\Delta_N)$, the formulas define $W_g>0$ and $\eta_{i\mid g}>0$, and plugging them back yields $w_i=W_{g(i)}\eta_{i\mid g(i)}$. Uniqueness on the interior follows because $W_g$ are uniquely recovered as sector sums, and then each $\eta^{(g)}$ is uniquely recovered by normalization.

\subsubsection{Definition: Block Covariance and Composite Sectors}
Let $\Sigma\in\mathbb{R}^{N\times N}$ be the (posterior) asset‑level covariance matrix, partitioned into blocks $\{\Sigma_{gh}\}_{g,h=1}^G$ with $\Sigma_{gh}\in\mathbb{R}^{n_g\times n_h}$.
Given within‑sector weights $\eta=\{\eta^{(g)}\}$, define the sector‑level (composite) covariance
\begin{equation}\label{eq:SigmaTilde}
\widetilde{\Sigma}=\widetilde{\Sigma}(\eta)\in\mathbb{R}^{G\times G},
\qquad
\widetilde{\Sigma}_{gh}\;=\;(\eta^{(g)})^\top \Sigma_{gh}\,\eta^{(h)}.
\end{equation}
Equivalently, if $R=(R_i)_{i\in\mathcal{I}}$ denotes asset returns and $R_g^\eta=\sum_{i\in\mathcal{I}_g}\eta_{i\mid g}R_i$ the composite sector returns, then $\mathrm{Cov}(R^\eta)=\widetilde{\Sigma}$.

\subsubsection{Definition: Risk Contributions}
For any $w\in\Delta_N$ define the portfolio variance $\sigma_p^2(w)=w^\top \Sigma w$; the marginal and total asset‑level risk contributions,
\[
\mathrm{MRC}_i(w) = 2(\Sigma w)_i,
\qquad
\mathrm{RC}_i(w) = w_i(\Sigma w)_i;
\]
and the sector‑level risk contribution $\mathrm{RC}_g(w)=\sum_{i\in\mathcal{I}_g}\mathrm{RC}_i(w)$.

Lemma: $\sum_{i=1}^N \mathrm{RC}_i(w)=\sigma_p^2(w)$ for all $w\in\Delta_N$. The proof is simple and follows that
$\sum_i w_i(\Sigma w)_i = w^\top \Sigma w$.

\subsubsection{Two‑level Variance and Sector Risk Decomposition}
Let $w=\Phi(W,\eta)$ be the final weights.
Then
\begin{equation}\label{eq:var_decomp}
\sigma_p^2(w) \;=\; W^\top \widetilde{\Sigma}(\eta)\, W,
\end{equation}
and, for each sector $g\in\mathcal{G}$,
\begin{equation}\label{eq:rc_sector}
\mathrm{RC}_g(w) \;=\; W_g\,\bigl(\widetilde{\Sigma}(\eta)\,W\bigr)_g.
\end{equation}

Proof:
Write the sector‑$g$ subvector of $w$ as $w^{(g)}=W_g\,\eta^{(g)}$. Then
\[
\sigma_p^2(w)
= \sum_{g,h} (w^{(g)})^\top \Sigma_{gh}\,w^{(h)}
= \sum_{g,h} W_g W_h\, (\eta^{(g)})^\top \Sigma_{gh}\,\eta^{(h)}
= W^\top \widetilde{\Sigma} W,
\]
which proves \eqref{eq:var_decomp}. For \eqref{eq:rc_sector},
\[
\mathrm{RC}_g(w)
= \sum_{i\in\mathcal{I}_g} w_i(\Sigma w)_i
= (w^{(g)})^\top \Bigl(\sum_h \Sigma_{gh}\,w^{(h)}\Bigr)
= \sum_h W_g W_h\, (\eta^{(g)})^\top \Sigma_{gh}\,\eta^{(h)}
= W_g\,(\widetilde{\Sigma} W)_g.
\]

This theorem leads to a corollary of Equivalence to Sector‑level Risk Parity that
Let $w=\Phi(W,\eta)$.
Then the condition ``$\mathrm{RC}_g(w)$ is equal for all $g$'' is equivalent to the (flat) risk‑parity condition for the $G$‑asset problem with weights $W$ and covariance $\widetilde{\Sigma}(\eta)$; i.e., $\mathrm{RC}^{\mathrm{sector}}_g(W)=W_g(\widetilde{\Sigma}W)_g$ is constant in $g$.The proof immediately follows from \eqref{eq:rc_sector}.

The within-sector equal risk characteristics demonstrated that if we fix $W$ and all $\eta^{(h)}$ for $h\neq g$.
Define the effective vector
\[
m^{(g)}(W,\eta)\;=\; W_g\,\Sigma_{gg}\,\eta^{(g)} \;+\; \sum_{h\neq g} W_h\,\Sigma_{gh}\,\eta^{(h)}\in\mathbb{R}^{n_g},
\]
which is the restriction of $(\Sigma w)$ to sector $g$.
For $i\in\mathcal{I}_g$ one has $\mathrm{RC}_i(w)=W_g\,\eta_{i\mid g}\,m^{(g)}_i(W,\eta)$.
Hence the within‑sector equal‑risk condition
$\mathrm{RC}_i(w)=\mathrm{RC}_j(w)$ for all $i,j\in\mathcal{I}_g$
is equivalent to the fixed‑point
\[
\eta^{(g)} \;\propto\; \frac{1}{\,m^{(g)}(W,\eta)\,}
\qquad\text{(componentwise reciprocal, followed by normalization onto }\Delta_{n_g}\text{)}.
\]
In the block‑diagonal special case $\Sigma_{gh}=0$ for $h\neq g$, this reduces to the classical within‑block risk‑parity condition $\eta^{(g)}\propto 1/(\Sigma_{gg}\eta^{(g)})$, which admits a unique solution on $\Delta_{n_g}$ when $\Sigma_{gg}\succ0$.
 
\subsubsection{Transaction Cost Model} Trading incurs proportional cost $c$ such that the transaction cost function for the entire trading period is defined by a linear function that $C_t(c,w) = c\sum_i |w_{i,t} - w_{i, t-1}|$

\begin{algorithm}[H]
\caption{Two-Level BHRP Fixed-Point Updates (Sector and Within-Sector)}
\label{alg:bhrp-fixedpoint}
\begin{algorithmic}[1]
\Require Posterior covariance $\Sigma\succ0$; sector partition $\{\mathcal{I}_g\}_{g=1}^G$; tolerance $\varepsilon>0$; max iters $K_{\max}$.
\Statex \textbf{Helpers:} $\textproc{Normalize}(x)=x/\mathbf{1}^\top x$;\quad
$\textproc{Block}(\Sigma,g,h)=\Sigma_{gh}$ (submatrix on $\mathcal{I}_g\times\mathcal{I}_h$)
\Statex \hspace{3.9em} $\textproc{RC}(w,\Sigma)=w\circ(\Sigma w)$; \quad $\textproc{RCVar}_{\text{within}}=\sum_g \mathrm{Var}\big(\mathrm{RC}_{\mathcal{I}_g}\big)$; \quad $\textproc{RCVar}_{\text{across}}=\mathrm{Var}\big(\sum_{i\in\mathcal{I}_g}\mathrm{RC}_i\big)$.
\State \textbf{Initialize:} $W^{(0)}\gets \textproc{Normalize}(\mathbf{1}_G)$; for each $g$, $\eta^{(g,0)}\gets \textproc{Normalize}(\mathbf{1}_{|\mathcal{I}_g|})$.
\For{$k=0,1,2,\dots,K_{\max}-1$}
  \State \textbf{Within-sector updates (Gauss--Seidel):}
  \For{$g=1$ to $G$}
     \State $\Sigma_{gg}\gets \textproc{Block}(\Sigma,g,g)$;\quad $m^{(g)}\gets W^{(k)}_g \Sigma_{gg}\eta^{(g,k)} + \sum_{h\neq g} W^{(k)}_h\, \textproc{Block}(\Sigma,g,h)\, \eta^{(h,k)}$
     \State $\eta^{(g,k+1)} \gets \textproc{Normalize}\big( (m^{(g)}\vee \delta)^{\odot(-1)} \big)$ \Comment{$\delta>0$ safeguards division; $\odot$ is componentwise power}
  \EndFor
  \State \textbf{Composite (sector) covariance:} for all $g,h$, $\widetilde{\Sigma}_{gh}\gets (\eta^{(g,k+1)})^\top \textproc{Block}(\Sigma,g,h)\, \eta^{(h,k+1)}$
  \State \textbf{Sector update (risk-parity fixed point):} $u\gets \widetilde{\Sigma}\,W^{(k)}$; \quad $W^{(k+1)} \gets \textproc{Normalize}\big((u\vee \delta)^{\odot(-1)}\big)$
  \State \textbf{Assemble weights:} for each $i\in\mathcal{I}_g$, $w^{(k+1)}_i \gets W^{(k+1)}_g \,\eta^{(g,k+1)}_{i\mid g}$
  \State \textbf{Stopping test:} 
        Compute $\mathrm{RC}\gets \textproc{RC}(w^{(k+1)},\Sigma)$; 
        $\Delta\gets \|w^{(k+1)}-w^{(k)}\|_1$;\;
        $\mathcal{E}\gets \textproc{RCVar}_{\text{within}}+\textproc{RCVar}_{\text{across}}$.
        \If{$\Delta \le \varepsilon$ \textbf{and} $\mathcal{E}\le \varepsilon$} \textbf{break} \EndIf
\EndFor
\Ensure Final weights $w^\star\gets w^{(k+1)}$ on $\Delta_N$.
\end{algorithmic}
\end{algorithm}

\begin{algorithm}[H]
\caption{Sector Risk-Parity via Log-Newton on $(W,\widetilde{\Sigma})$}
\label{alg:sector-lognewton}
\begin{algorithmic}[1]
\Require $\widetilde{\Sigma}\succ0$; budgets $b\in\mathbb{R}^G_{++}$ with $\mathbf{1}^\top b=1$; tolerance $\varepsilon$.
\State Initialize $y\gets \mathbf{0}$ (so $u=\exp(y)=\mathbf{1}$), and enforce simplex by $g(y)=\mathbf{1}^\top \exp(y)-1=0$.
\Repeat
   \State $u\gets \exp(y)$;\quad $z\gets \widetilde{\Sigma}u$;\quad $U\gets \mathrm{diag}(u)$;\quad $Z\gets \mathrm{diag}(z)$
   \State Residuals: $r\gets \log z + y - \log b - c\,\mathbf{1}$ with scalar $c$ s.t.\ $g(y)=0$;\quad $s\gets g(y)$
   \State Jacobian: $J\gets I + Z^{-1}\widetilde{\Sigma}U$;\quad Solve
          $\begin{bmatrix}J & -\mathbf{1}\\ \mathbf{1}^\top U & 0\end{bmatrix}
           \begin{bmatrix}\Delta y\\ \Delta c\end{bmatrix}
           = -\begin{bmatrix}r\\ s\end{bmatrix}$
   \State Line search $\alpha\in(0,1]$ to decrease $\|r\|_2^2+s^2$;\quad $y\gets y+\alpha\Delta y$;\quad $c\gets c+\alpha\Delta c$
\Until{$\|r\|_\infty \le \varepsilon$ and $|s|\le \varepsilon$}
\Ensure $W^\star=\exp(y)$ (already on $\Delta_G$).
\end{algorithmic}
\end{algorithm}

\begin{algorithm}[H]
\caption{Sequential Bayesian Hierarchical Update of $(\widehat{\mu},\Sigma)$}
\label{alg:bayes-filter}
\begin{algorithmic}[1]
\Require New returns $\{R_{i,t}\}_{i=1}^N$; rolling window $\mathcal{W}_t$; sector map $g(\cdot)$; prior scales $\{\tau_i^2\}$.
\State Compute per-asset sample mean/variance on $\mathcal{W}_t$: $\bar r_{i}$ and $s_i^2$; let $n=|\mathcal{W}_t|$.
\State Sector priors: $\mu_g \gets \frac{1}{|\mathcal{I}_g|}\sum_{i\in\mathcal{I}_g}\bar r_i$.
\State Posterior means (precision‑weighted): $\widehat{\mu}_i \gets \dfrac{\tau_i^{-2}\mu_{g(i)} + n\,s_i^{-2}\bar r_i}{\tau_i^{-2} + n\,s_i^{-2}}$.
\State Posterior covariance $\Sigma \gets \text{Ledoit--Wolf}(\{R_{i,s}\}_{s\in\mathcal{W}_t})$ (or other shrinkage).
\Ensure Updated $(\widehat{\mu},\Sigma)$ for use in Algorithms~\ref{alg:bhrp-fixedpoint}–\ref{alg:sector-lognewton}.
\end{algorithmic}
\end{algorithm}

\subsection{Reinforcement Learning Layer}

Trading occurs on a discrete rebalance grid $t=0,1,2,\dots$.
At each $t$ we observe returns up to $t$ and update a Bayesian hierarchical filter to obtain the posterior mean vector $\widehat{\mu}_t$ and covariance matrix $\Sigma_t\succ0$ for one‑period‑ahead returns. We then select portfolio weights $w_t$ that are implemented over $[t,t{+}1]$.

\subsubsection{MDP specification}
MDP is a common technique in reinforcement learning. In our reinforcement learning layer, we define the following: 
\subsubsection{Definition: State, Action, Transition, Reward}
The portfolio control problem is cast as an average‑reward MDP $\mathcal{M}=(\mathcal{S},\mathcal{A},P,U)$:
\begin{itemize}
  \item \emph{State} $s_t\in\mathcal{S}\subset\mathbb{R}^d$ collects $k$ lags of period returns, the posterior summaries $(\widehat{\mu}_t,\mathrm{diag}(\Sigma_t))$, and the previous weights $w_{t-1}$. The state space is compact because returns are bounded and weights lie on a simplex.
  \item \emph{Action} $a_t$ is a two‑level weight parametrization (Definition~\ref{def:twolevel}) with sector weights $W_t\in\Delta_G$ and conditional within‑sector weights $\eta_t^{(g)}\in\Delta_{n_g}$; the final asset weights are $w_{i,t}=W_{g(i),t}\,\eta_{i\mid g(i),t}$.
  \item \emph{Transition} $P(\cdot|s_t,a_t)$ is induced by the one‑step return realization $R_{t+1}$ and the Bayesian update from data at $t{+}1$, producing $(\widehat{\mu}_{t+1},\Sigma_{t+1})$.
  \item \emph{Reward} at $t{+}1$ is
  \begin{equation}\label{eq:reward}
    U_{t+1}(w_t) \;=\; w_t^\top R_{t+1}
    \;-\; c\,\|w_t-w_{t-1}\|_1
    \;-\; \lambda\Bigl[\alpha\,\mathcal{V}_{\mathrm{within}}(w_t;\Sigma_t) + (1{-}\alpha)\,\mathcal{V}_{\mathrm{across}}(w_t;\Sigma_t)\Bigr],
  \end{equation}
  where $c>0$ is proportional transaction cost, $\lambda>0$ tunes risk‑budget tightness, and $\alpha\in[0,1]$ splits within/ across‑sector penalties defined below.
\end{itemize}

\subsubsection{Definition: Hierarchical risk‑parity penalty}
Let $\mathrm{RC}(w_t)\in\mathbb{R}^N$ be the vector of asset‑level total risk contributions
$\mathrm{RC}_i(w_t)=w_{i,t}\,(\Sigma_t w_t)_i$ and define sector aggregation by $S\in\{0,1\}^{G\times N}$ with $S_{g,i}=\mathbf{1}\{i\in\mathcal{I}_g\}$, so $\mathrm{RC}^{\mathrm{sec}}(w_t)=S\,\mathrm{RC}(w_t)$.
Let $P_N=I_N-\frac{1}{N}\mathbf{1}\mathbf{1}^\top$ and $P_G=I_G-\frac{1}{G}\mathbf{1}\mathbf{1}^\top$ denote centering operators.
We measure dispersion by (scaled) squared norms:
\[
\mathcal{V}_{\mathrm{within}}(w_t;\Sigma_t)=\frac{1}{N}\,\bigl\|P_N\,\mathrm{RC}(w_t)\bigr\|_2^2,
\qquad
\mathcal{V}_{\mathrm{across}}(w_t;\Sigma_t)=\frac{1}{G}\,\bigl\|P_G\,S\,\mathrm{RC}(w_t)\bigr\|_2^2.
\]
Both vanish iff asset‑level (respectively sector‑level) risk contributions are equal.

\subsubsection{Policy Parametrization and Feasibility}

We use a factorized softmax policy that maps logits to simplexes.

We define a Factorized Softmax Policy as given state $s$, the actor outputs logits $\Psi_g(s)\in\mathbb{R}$ for sectors and $\Phi_{i\mid g}(s)\in\mathbb{R}$ within each sector.
The corresponding probabilities are
\[
W_g(s)=\frac{\exp(\Psi_g(s))}{\sum_{h=1}^G \exp(\Psi_h(s))},
\qquad
\eta_{i\mid g}(s)=\frac{\exp(\Phi_{i\mid g}(s))}{\sum_{j\in\mathcal{I}_g}\exp(\Phi_{j\mid g}(s))}.
\]
The final weights are $w_{i}(s)=W_{g(i)}(s)\,\eta_{i\mid g(i)}(s)$.

\subsubsection{Definition: Feasibility and Representability}
For every state $s$, the map $(\Psi,\Phi)\mapsto w(s)$ yields $w(s)\in\Delta_N$. Moreover, for any $w\in\mathrm{int}(\Delta_N)$ there exist logits producing exactly $w$; e.g., choose
$\Psi_g=\log\!\Bigl(\sum_{i\in\mathcal{I}_g}w_i\Bigr)+c$ and
$\Phi_{i\mid g}=\log\!\Bigl(w_i/\sum_{j\in\mathcal{I}_g}w_j\Bigr)$.

Let $w=w(s)$ and write $g(i)$ for the sector of asset $i$.
For any $g$ and $j\in\mathcal{I}_h$,
\[
\frac{\partial w_i}{\partial \Psi_g}=w_i\bigl(\mathbf{1}\{g(i)=g\}-W_g\bigr),
\qquad
\frac{\partial w_i}{\partial \Phi_{j\mid h}}=W_{g(i)}\,\eta_{i\mid g(i)}\bigl(\mathbf{1}\{i=j\}-\eta_{j\mid h}\bigr).
\]

\subsubsection{Reward Regularity and Policy‑gradient Applicability}

We impose standard regularity:
(i) returns are bounded almost surely, $\lvert R_{i,t}\rvert\le L_R$ for all $i,t$;
(ii) $\Sigma_t$ is symmetric positive definite and $\|\Sigma_t\|_2\le L_\Sigma$; (iii) the state space is compact (finite lags of bounded variables) and the action space is $\Delta_G\times\prod_g\Delta_{n_g}$.

Lemma: Boundedness and Lipschitz continuity
Under the regularity above, the reward $U_{t+1}(w)$ in \eqref{eq:reward} is bounded and globally Lipschitz in $w$ on $\Delta_N$, with a constant depending only on $(L_R,L_\Sigma,c,\lambda)$.
\\
Proof:
The linear term $w^\top R_{t+1}$ is bounded by $L_R$. The cost term is $c\|w-w_{t-1}\|_1\le 2c$ on the simplex and is $1$‑Lipschitz. For the penalty, $\mathrm{RC}(w)=w\circ (\Sigma_t w)$, so its Jacobian is $J_{\mathrm{RC}}(w)=\mathrm{diag}(\Sigma_t w)+\mathrm{diag}(w)\Sigma_t$; thus $\|J_{\mathrm{RC}}(w)\|_2\le \|\Sigma_t\|_2+\|\Sigma_t\|_2\le 2L_\Sigma$. The maps $w\mapsto P_N\mathrm{RC}(w)$ and $w\mapsto P_G S \mathrm{RC}(w)$ are linear post‑compositions of $\mathrm{RC}$ with operator norms $\le 1$. The squared norms are $C^1$ with gradient bounded by $2\|(P\cdot)^\top\|\cdot\|\mathrm{RC}(w)\|\,\|J_{\mathrm{RC}}(w)\|$, which is finite on the compact simplex. Summing yields the claim.

Theorem: Policy‑gradient identity and existence of optimal stationary policies
Consider the average‑reward objective $J(\theta)=\liminf_{T\to\infty}\frac{1}{T}\,\mathbb{E}_\theta\bigl[\sum_{t=0}^{T-1}U_{t+1}\bigr]$ under a differentiable stochastic policy $\pi_\theta$ defined by the factorized softmax above. If the induced Markov chain on $\mathcal{S}$ under any stationary policy is irreducible and positive recurrent, then:
(i) an optimal stationary policy exists; and
(ii) the gradient satisfies the policy‑gradient theorem
\[
\nabla_\theta J(\theta)=\mathbb{E}_\theta\bigl[\nabla_\theta \log \pi_\theta(a_t\mid s_t)\,\cdot\,Q^{\pi_\theta}(s_t,a_t)\bigr],
\]
where $Q^{\pi_\theta}$ is the action‑value function. The expectation exists and is finite by the Lemma

Proof: 
Existence follows from average‑reward MDP theory on compact state–action spaces with bounded rewards. The policy‑gradient identity is standard for differentiable policies when rewards are integrable and the chain is unichain; boundedness and Lipschitzness ensure the interchange of gradient and expectation is valid. The detailed proof is in the appendix section. 

\subsubsection{Closed‑form gradients for the RP penalty}

Let $\mathrm{RC}(w)=w\circ(\Sigma_t w)$ and write $J_{\mathrm{RC}}(w)=\mathrm{diag}(\Sigma_t w)+\mathrm{diag}(w)\Sigma_t$.
Define $P_N=I-\frac{1}{N}\mathbf{1}\mathbf{1}^\top$ and $P_G$ analogously.
Then
\[
\nabla_w \,\mathcal{V}_{\mathrm{within}}(w;\Sigma_t)
= \frac{2}{N}\,J_{\mathrm{RC}}(w)^\top P_N\,\mathrm{RC}(w),
\qquad
\nabla_w \,\mathcal{V}_{\mathrm{across}}(w;\Sigma_t)
= \frac{2}{G}\,J_{\mathrm{RC}}(w)^\top S^\top P_G\,S\,\mathrm{RC}(w).
\]
By chain rule, $\nabla_\theta U_{t+1} = \bigl(\nabla_w U_{t+1}\bigr)\,\bigl(\nabla_\theta w_\theta(s_t)\bigr)$ with $\nabla_\theta w_\theta$ obtained from the softmax derivatives above.

\subsubsection{Algorithm: PPO with factorized actions}

We use Proximal Policy Optimization (PPO) with the factorized actor and a baseline critic.

\begin{algorithm}[H]
\caption{PPO for RL--BHRP}\label{alg:ppo-rl-bhrp}
\begin{algorithmic}[1]
\Require Posterior updater for $(\widehat{\mu}_t,\Sigma_t)$; penalty $(\lambda,\alpha)$; cost $c$; PPO hyperparameters
\State Initialize actor--critic $(\theta,\phi)$; set $w_{-1}\gets$ equal-weight
\For{iteration $k=1,2,\dots$}
  \State Roll out $T$ steps: observe $s_t$, sample $a_t\sim\pi_\theta(\cdot\mid s_t)$, map to $w_t$, trade, observe $U_{t+1}$, update posterior
  \State Compute advantages $\widehat{A}_t$ (e.g., GAE) and value targets
  \State Update $\theta$ via clipped surrogate with entropy bonus; update critic $\phi$
\EndFor
\State \Return trained policy $s\mapsto (W(s),\eta(s))$
\end{algorithmic}
\end{algorithm}

\subsubsection{Optional projection to exact hierarchical risk parity}

If exact equality of risk contributions is required at trade time, add a post‑processing projection
$w_t \leftarrow \arg\min_{w\in\Delta_N}\|w-w_t\|_2^2$ subject to asset‑ and sector‑level equal‑risk constraints.
A practical alternative is a two‑stage fixed‑point: (i) solve sector RP for $(W_t,\widetilde{\Sigma}(\eta_t))$; (ii) solve within‑sector RP in each block $\Sigma_{gg}$; iterate a few steps.

\section{Results}
Due to computational constraints, we evaluate the proposed RL‑BHRP framework on a rules‑based subset of the investable universe defined in the methodology. We obtain adjusted close (total-return) prices from Yahoo Finance from January 1, 2012, to August 13, 2025. The sample is partitioned into a training period (January 1, 2012, to January 1, 2020) used for model development and hyperparameter calibration, and a testing period (January 1, 2020, to August 31, 2025) reserved strictly for out-of-sample evaluation. Because the daily data end on 2025-08-13, the final calendar month is partial. Our last monthly return reflects performance from the month-end to August 13, 2025, and annualizations use the standard 12-month convention for monthly data. \\
\\
Universe construction adheres to the screening and timing rules outlined in the methodology: membership is determined ex ante at each quarter-end using liquidity (20-day ADV) and price thresholds, and then frozen intra-quarter to prevent look-ahead. At each monthly rebalance, assets without a well‑defined period return are masked, and remaining weights are renormalized on the simplex. Within this protocol we select a sector‑balanced subset of large‑ and mega‑cap names spanning all GICS sectors (Table \ref{tab:sector-map-etf-subset}). This preserves the cross-sectional breadth of the hierarchical prior and within-sector risk parity while keeping the computational budget tractable.

\begin{table}[H]
\centering
\caption{Sector map for the evaluation subset with corresponding ETFs}
\label{tab:sector-map-etf-subset}
\setlength{\tabcolsep}{6pt}
\renewcommand{\arraystretch}{1.1}
\begin{tabular}{
  >{\raggedright\arraybackslash}p{0.26\linewidth}
  >{\raggedright\arraybackslash}p{0.10\linewidth}
  >{\raggedright\arraybackslash}p{0.58\linewidth}}
\toprule
\textbf{Sector} & \textbf{ETF Ticker} & \textbf{Constituents (tickers)} \\
\midrule
Information Technology & XLK & AAPL, MSFT, NVDA, ADBE, ORCL, CSCO, INTC, AVGO, AMD, AMAT \\
Communication Services & XLC & GOOGL, META, T, VZ \\
Consumer Discretionary & XLY & AMZN, HD, NKE, LOW \\
Consumer Staples & XLP & PG, KO, PEP, COST, PM \\
Financials & XLF & JPM, BAC, WFC, V, MA \\
Energy & XLE & XOM, CVX \\
Health Care & XLV & LLY, UNH, ABBV, TMO, PFE, MRK \\
Industrials & XLI & CAT, BA, MMM, GE, HON \\
Utilities & XLU & DUK, SO, D \\
Real Estate & XLRE & PLD, SPG \\
Materials & XLB & LIN, APD \\
\bottomrule
\end{tabular}
\end{table}

\subsection{Cumulative Performance}
\begin{figure}[H]
\centering
\captionsetup{font=small} 
\setlength{\tabcolsep}{4pt}

\begin{minipage}[t]{0.56\linewidth}
\vspace{0pt} 
\includegraphics[width=\linewidth]{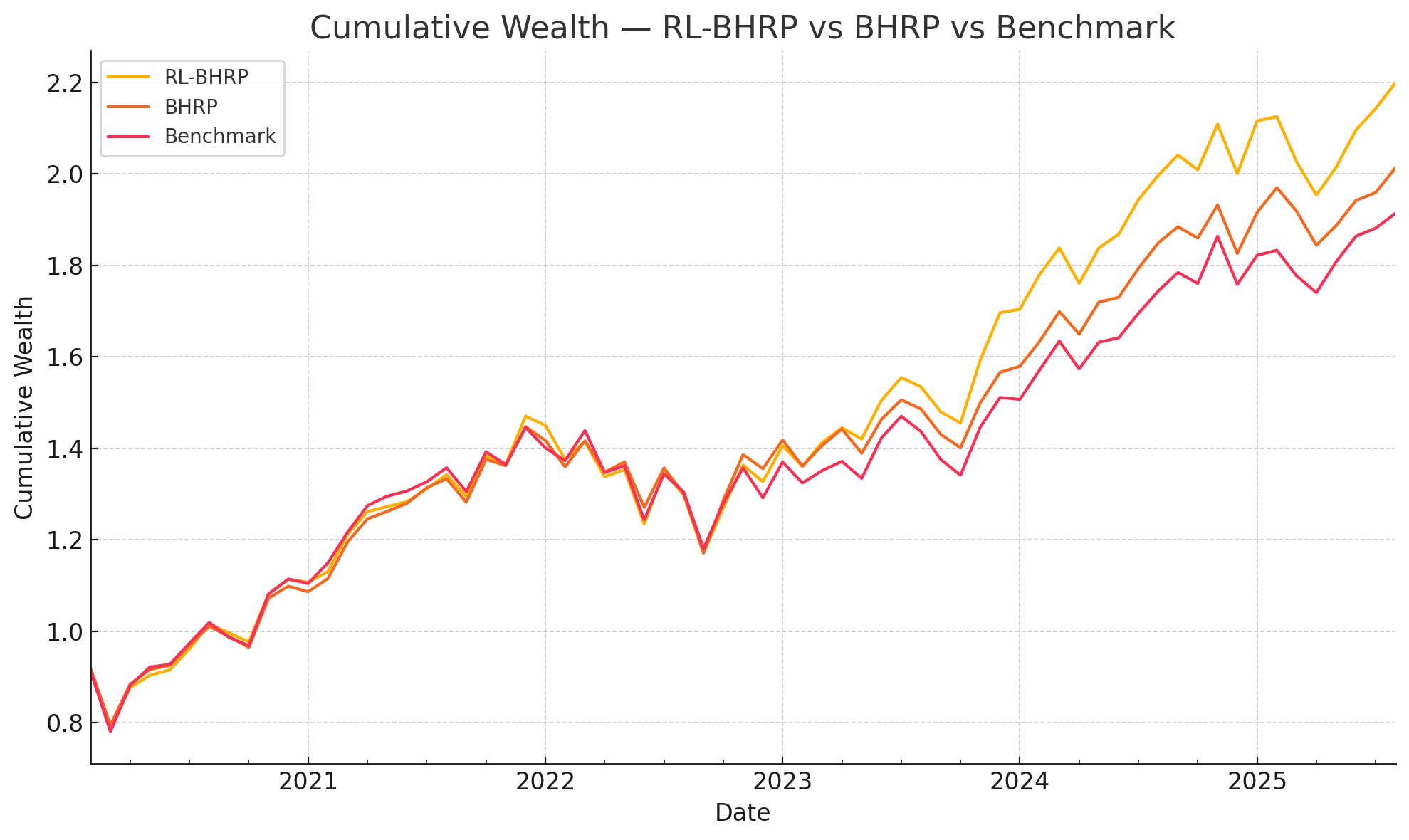}
\captionof{figure}{Testing Period Cumulative Returns}
\label{fig:cum_wealth}
\end{minipage}
\hfill
\begin{minipage}[t]{0.41\linewidth}
\vspace{0pt} 
\captionof{table}{Full-Period Metrics}
\label{tab:table1}
\scriptsize
\begin{tabular}{lccc}
\hline
\textbf{Metric} & \textbf{RL-BHRP} & \textbf{BHRP} & \textbf{Benchmark} \\
\hline
Periods & 67 & 67 & 67 \\
Start & 2020-02-29 & 2020-02-29 & 2020-02-29 \\
End & 2025-08-31 & 2025-08-31 & 2025-08-31 \\
Total Cumulative Return & 1.199612 & 1.013672 & 0.914063 \\
CAGR (Geometric Ann.) & 0.151637 & 0.133563 & 0.123310 \\
Annual Return (Mean$\times$12) & 0.157213 & 0.139742 & 0.131883 \\
Annual Volatility & 0.173748 & 0.165164 & 0.172838 \\
Sharpe & 0.904829 & 0.846076 & 0.763042 \\
Sortino & 1.648382 & 1.534682 & 1.368210 \\
Max Drawdown & -0.203308 & -0.191023 & -0.183210 \\
Calmar & 0.745850 & 0.699201 & 0.673053 \\
Tracking Error & 0.036883 & 0.035790 & 0.000000 \\
Information Ratio & 0.686752 & 0.219576 & NaN \\
Beta & 0.982509 & 0.935146 & 1.000000 \\
Jensen Alpha & 0.027636 & 0.016412 & 0.000000 \\
VaR 5\% (period) & -0.074066 & -0.067241 & -0.079989 \\
CVaR 5\% (period) & -0.102409 & -0.097049 & -0.103351 \\
Hit Rate ($>$0) & 0.641791 & 0.641791 & 0.626866 \\
\hline
\end{tabular}
\end{minipage}
\end{figure}

Our primary benchmark is an equal‑weight sector portfolio composed of the SPDR sector ETFs (XLC, XLY, XLP, XLE, XLF, XLV, XLI, XLK, XLB, XLRE, XLU). Benchmark returns are computed from adjusted closes, resampled at month‑end, and rebalanced monthly to equal weights. All strategy returns are evaluated on the same calendar grid, with annualization based on a 12‑period year. Transaction costs for RL‑BHRP follow the paper’s cost model; the benchmark is cost‑free by construction. In the Results section we report (i) cumulative wealth curves for RL‑BHRP vs. BHRP vs. Benchmark, (ii) full‑period and year‑by‑year performance tables (CAGR, volatility, Sharpe, Sortino, max drawdown, tracking error, information ratio, CAPM alpha/beta), and (iii) dynamic weight diagnostics to document portfolio adaptivity over time.\\
\\
Figure \ref{fig:cum_wealth} shows that RL-BHRP compounds to the highest terminal wealth during the out-of-sample window (Feb - 2020 to Aug - 2020), with a visibly wider gap from mid-2023 onward. Table \ref{tab:table1} confirms the dominance in levels and on a risk‑adjusted basis. RL-BHRP earns a total cumulative return of 1.1996 ($\sim$ 120\%), versus 1.0137 for BHRP and 0.9141 for the sector benchmark, implying a CAGR of 15.16\%, which is +1.81 pp above BHRP and +2.83 pp above the benchmark. Despite slightly higher volatility than BHRP (17.37\% vs 16.52\%; benchmark 17.28\%), RL-BHRP posts the highest Sharpe (0.905) and Sortino (1.648), indicating that the incremental return more than compensates for the modest rise in risk. Drawdowns are comparable across strategies; RL-BHRP’s trough ($\sim$ 20.33\%) is marginally deeper than BHRP ($\sim$ 19.10\%) and the benchmark ($\sim$ 18.32\%), yet its Calmar (0.746) exceeds both alternatives because of the higher geometric growth.\\
\\
RL--BHRP remains benchmark-proximal (beta \(0.983\); tracking error \(3.69\%\)) yet consistently adds value: the information ratio is \(0.687\) (BHRP: \(0.220\)), and the Jensen alpha is \(2.76\%\) (BHRP: \(1.64\%\)). The hit rate (fraction of months with positive returns) is also favorable for RL--BHRP at \(0.642\) (benchmark \(0.627\)). Tail-risk measures indicate the expected trade-off for an adaptive allocator: period \(\mathrm{VaR}_{5\%}\) and \(\mathrm{CVaR}_{5\%}\) for RL--BHRP are \(-7.41\%\) and \(-10.24\%\), respectively, lying between BHRP (less severe: \(-6.72\%\), \(-9.70\%\)) and the benchmark (more severe: \(-8.00\%\), \(-10.34\%\)). Taken together, these outcomes align with the design of RL--BHRP: Bayesian shrinkage stabilizes inputs; the hierarchical risk-parity structure limits concentration and sector imbalances; and the policy-gradient layer reallocates toward time-varying opportunities without materially increasing active risk. The result is a meaningful improvement in both absolute and benchmark-relative performance---higher CAGR, Sharpe, and alpha---achieved with low tracking error and only a modest increase in downside tail measures relative to the static BHRP baseline.

\subsection{Weights}
\begin{figure}[H]
\centering
\begin{minipage}[t]{0.48\linewidth}
    \centering
    \includegraphics[width=\linewidth]{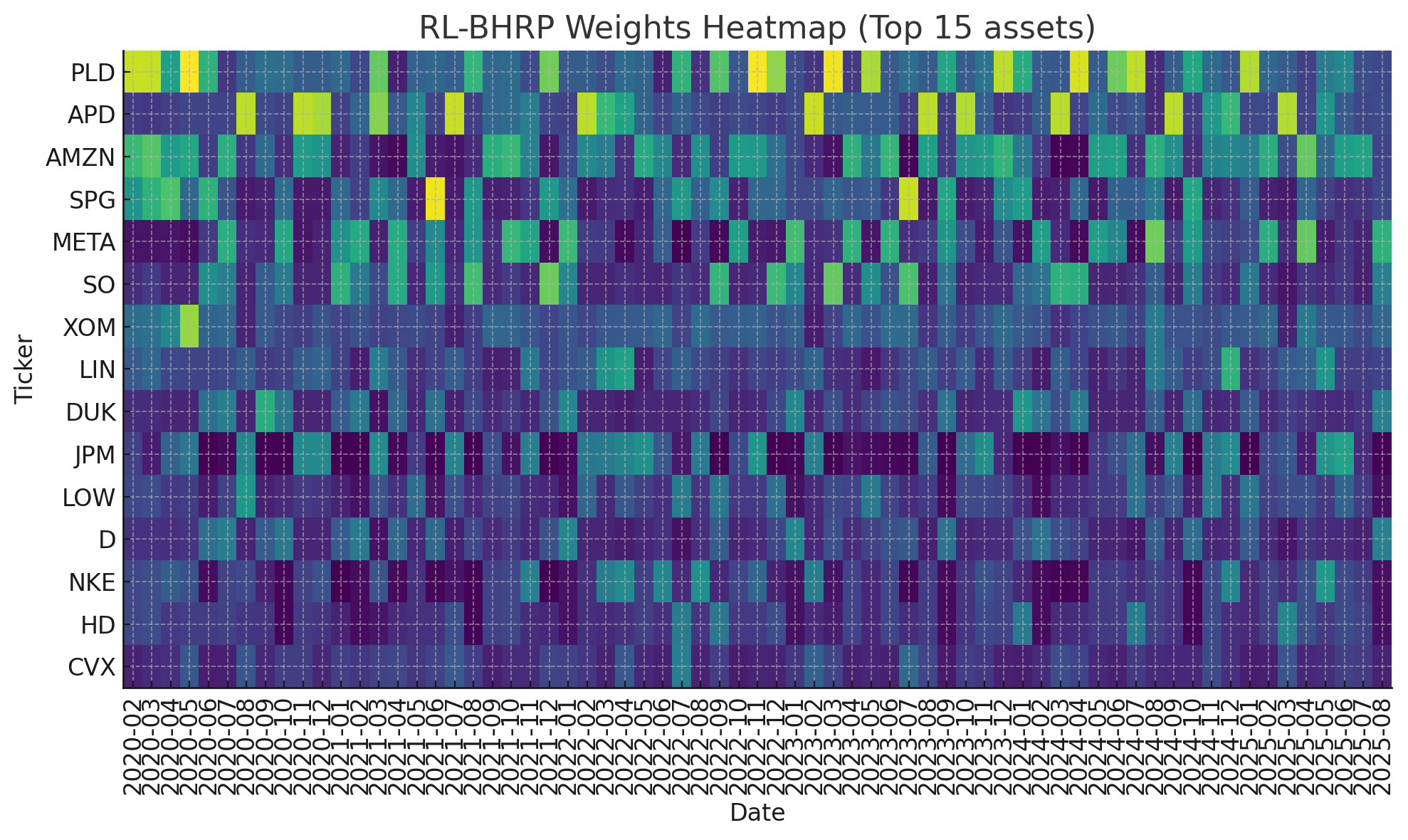}
    \caption{Weights Heatmap}
    \label{fig:weights1}
\end{minipage}
\hfill
\begin{minipage}[t]{0.48\linewidth}
    \centering
    \includegraphics[width=\linewidth]{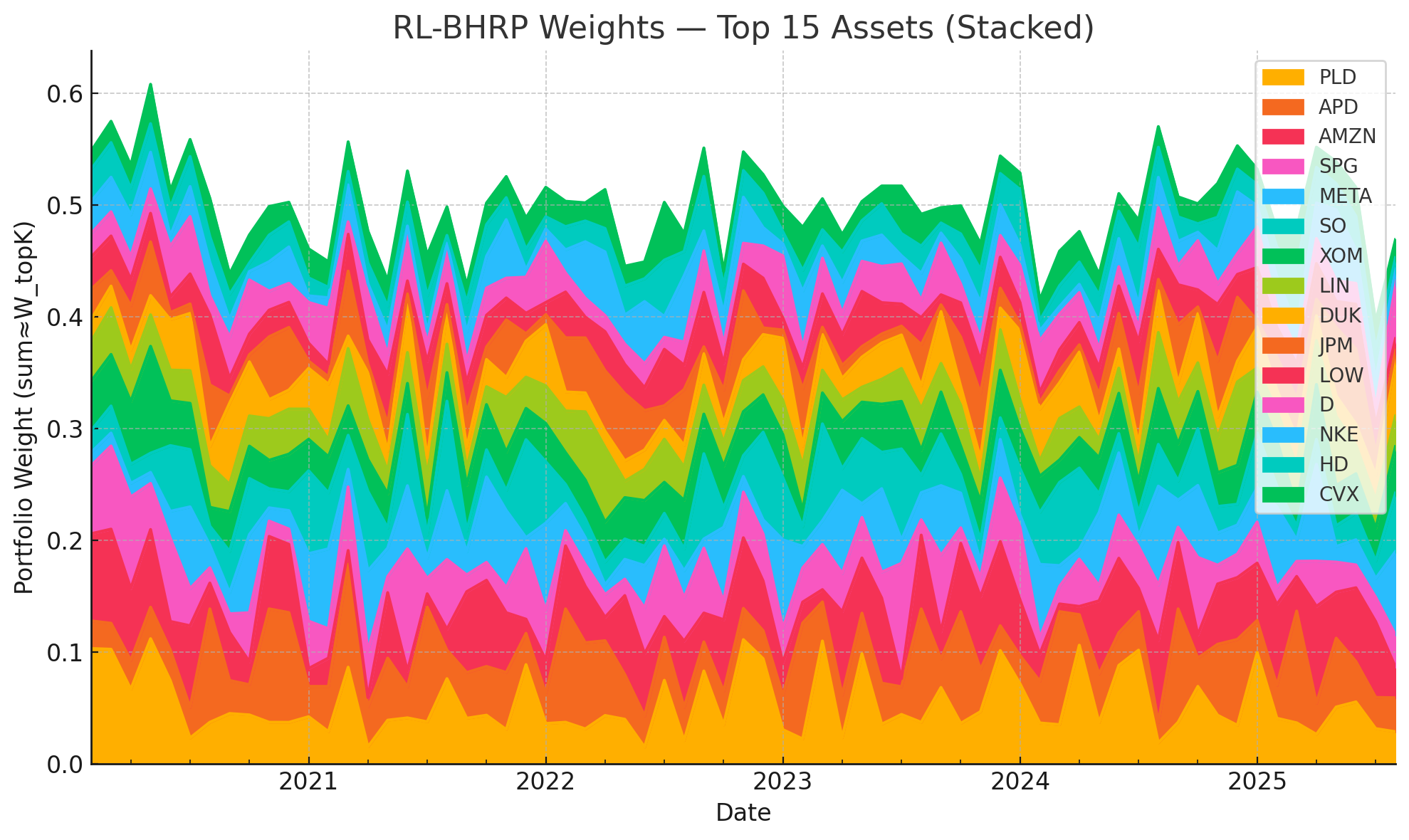}
    \caption{EL-BHRP Dynamic Weights Stack}
    \label{fig:weights2}
\end{minipage}
\end{figure}

Figures \ref{fig:weights1} and \ref{fig:weights2} summarize the evolution of RL--BHRP portfolio weights over the out-of-sample period. The heatmap (Figure \ref{fig:weights1}) reveals \emph{structured but adaptive} allocations: intensity varies over time across a diversified set of large and liquid names from multiple sectors (e.g., \texttt{PLD}, \texttt{APD}, \texttt{AMZN}, \texttt{META}, \texttt{SO}, \texttt{XOM}, \texttt{LIN}, \texttt{DUK}, \texttt{JPM}, \texttt{LOW}, \texttt{D}, \texttt{NKE}, \texttt{HD}, \texttt{CVX}). No single position persists at a dominating level; instead, exposures rotate gradually as posterior signals and the hierarchical risk-parity penalty evolve. The stacked area in Figure \ref{fig:weights2} shows that the aggregate share of the Top 15 names (selected ex post by average weight) remains bounded and varies smoothly, which is consistent with the dispersion penalty in the reward discouraging excessive concentration and with the transaction-cost term limiting needless turnover.\\
\\ 
Let \(w_t\in\Delta_N\) denote the asset-weight vector at month-end \(t\).
Define the Top--\(K\) set \(\mathcal{T}\) as the \(K=15\) assets with the largest time-average weights over the test window, and set
\[
W_{\text{Top15}}(t)\;=\;\sum_{i\in\mathcal{T}} w_{i,t}\!,
\qquad
\tau_t\;=\;\lVert w_t-w_{t-1}\rVert_1,
\qquad
\rho_t\;=\;\frac{w_t^\top w_{t-1}}{\lVert w_t\rVert_2\,\lVert w_{t-1}\rVert_2}.
\]
Here \(W_{\text{Top15}}(t)\) measures concentration in the leading names, \(\tau_t\) is L1 turnover, and \(\rho_t\) is a unit-free measure of \emph{persistence} (cosine similarity) in allocations. Over the out-of-sample window we report their time-averages \(\overline{W}_{\text{Top15}}=\frac{1}{T}\sum_t W_{\text{Top15}}(t)\), \(\overline{\tau}=\frac{1}{T}\sum_t \tau_t\), and \(\overline{\rho}=\frac{1}{T}\sum_t \rho_t\). \\
\\
Using the saved RL weights, the portfolio traded \(N\!=\!48\) distinct names. The \emph{Top--15} share averaged
\(\overline{W}_{\text{Top15}}\approx 0.50\) (median \(\approx 0.50\), range \(\approx[0.40,0.61]\)),
indicating that roughly half of the capital was allocated to the most influential names while the remaining half was diversified across the long tail. Turnover averaged \(\overline{\tau}\approx 0.60\) (95th percentile \(\approx 0.85\)), which, under the 5~bps one-way cost model, implies average frictions on the order of \(\approx 3\)~bps per month---two orders of magnitude smaller than the strategy's average monthly return. The mean cosine similarity \(\overline{\rho}\approx 0.74\) confirms high month-to-month persistence with gradual tilts rather than abrupt flips. Visually, the heatmap exhibits banded, slowly varying intensity by name, and the stacked chart shows migration of weight across sectors (e.g., periodic increases among Energy and Utilities names alongside shifts in Consumer and Technology/Communication holdings) without violating the risk-budget constraints. \\
\\ 
These diagnostics are consistent with the design of RL--BHRP. The Bayesian layer stabilizes inputs; the two-level risk-parity structure prevents single-name or single-sector dominance; and the policy-gradient layer reallocates toward time-varying opportunities with smooth adjustments that respect the cost term. The result is an adaptive, diversified portfolio in which concentration is actively bounded, sector balance is preserved, and rebalancing dynamics remain economically plausible rather than reactive or noisy, as evidenced by the heatmap and stacked-weight trajectories in Figures \ref{fig:weights1} and \ref{fig:weights2}.

\section{Limitations and Future Research}

\subsection{Data and Universe Construction}
The empirical analysis relies on Yahoo Finance adjusted closes. While convenient and widely used, this source can omit delistings and occasionally mis-handle corporate actions, which introduces survivorship risk and measurement noise that may bias performance upward. To preserve computational tractability, we further evaluate a sector-balanced \emph{subset} of the mechanically screened tradable universe; although the quarterly screen is ex-ante and frozen intra-quarter, the reduction in cross-sectional breadth can attenuate the benefits of hierarchical shrinkage and within-sector risk parity. Replicating the study on survivorship-free, point-in-time data (e.g., CRSP/Refinitiv/Compustat with delisting returns and historical identifiers), and scaling to the full screened universe, such as the S\&P500 or Russell1000, would strengthen external validity. Finally, because the last calendar month of our sample is partial, annualized statistics are slightly perturbed; robustness checks should either truncate that month or enforce a strict end-of-month grid for all assets.

\subsection{Modeling Restrictions and Priors}
The Bayesian layer uses Normal likelihoods for returns with Ledoit--Wolf shrinkage for covariances. Equity returns, however, display heavy tails, skewness, and volatility clustering that are imperfectly captured by Gaussian assumptions. A more faithful specification would substitute heavy-tailed or skewed likelihoods (e.g., Student-$t$ or skew-$t$), robust scatter estimators (e.g., Tyler’s M-estimator), or dynamic structures such as EWMA/DCC, factor stochastic volatility, or state-space hierarchies. In addition, the current empirical-Bayes treatment fixes prior scales exogenously; learning hyperpriors from data---for example via hierarchical horseshoe priors or marginal likelihood maximization---could improve shrinkage in thinly observed names and adapt priors to regime changes.

\subsection{Risk Budgeting and Portfolio Constraints}
The hierarchical risk-parity component is enforced through a quadratic dispersion penalty on asset- and sector-level risk contributions. This soft penalty encourages, but does not guarantee, exact equality, and its quadratic form may overweight large deviations relative to small, systematic ones. Post-trade projection to the equal-risk manifold, or alternative divergences based on entropy or absolute deviations, offer principled ways to tighten risk budgets. Our implementation also assumes long-only, fully invested portfolios with monthly rebalancing and proportional $L^1$ transaction costs. Realistic mandates often impose leverage bounds, short-sale constraints, turnover budgets, minimum-lot sizes, and sector or name caps. Embedding these as hard constraints or barrier terms, together with impact-aware costs (e.g., square-root impact scaled by volatility and ADV), would enhance implementability and capacity analysis.

\subsection{RL Design and Training}
We use PPO with a factorized softmax over sector and within-sector logits. Although robust and interpretable, PPO can be sensitive to reward scaling and hyperparameters, and it optimizes expected return rather than explicit downside criteria. Risk-sensitive alternatives---distributional RL, CVaR policy gradients, or constrained actor--critic with Lagrangian updates---may deliver tighter control of tail risk while preserving performance. The current state vector aggregates lagged returns and posterior summaries; it omits macro covariates, option-implied measures, and learned regime indicators, and it models temporal dependence only through finite lags. Recurrent encoders (GRU/LSTM), attention over longer contexts, or learned latent states (HMM/SSM) could mitigate partial observability. Finally, the policy is trained on 2012--2019 and evaluated post-2020 without re-training; distribution shift can erode performance. Rolling or expanding-window re-training with strict nesting, conservative offline RL, and robustness to covariance shocks would address nonstationarity explicitly.

\subsection{Evaluation and Inference}
Our design adopts a single anchored train/test split. To support stronger inferential claims, future evaluations should include rolling-origin tests across multiple non-overlapping windows, factor-model alphas (e.g., FF5+Momentum) with Newey--West standard errors, and formal tests of performance differences such as Ledoit--Wolf or Jobson--Korkie tests for Sharpe ratios. When comparing many model variants, the Superior Predictive Ability and Model Confidence Set procedures are advisable. Beyond statistical significance, economic significance should be stress-tested by varying cost assumptions (0--25~bps one way), scaling costs by volatility and liquidity, and quantifying capacity under impact-aware execution.

\subsection{Theoretical Aspects}
While we establish the feasibility of the two-level weight construction and state a policy-gradient identity under standard regularity, several theoretical questions remain open. The existence, uniqueness, and convergence rates of the hierarchical RP fixed point under realistic block structures have not been derived; sufficient conditions based on block diagonal dominance or restricted eigenvalue bounds warrant further investigation. For the RL layer, generalization guarantees under nonstationarity are not available; deriving regret or stability bounds for average-reward RL with Lipschitz rewards and mixing conditions would provide principled guidance on sample complexity and robustness.

\subsection{Future Research}
A coherent next step is to replicate all results on survivorship-free data with point-in-time classification, extend the universe to broad indices in the U.S. and internationally, and compare monthly, weekly, and daily rebalancing under impact-aware costs. On the modeling side, embedding heavy-tailed/dynamic covariances within the Bayesian hierarchy and learning hyperpriors endogenously should be paired with risk-sensitive RL objectives and constrained actor--critic algorithms. At the interface of learning and optimization, unrolling the BHRP fixed point as a differentiable layer would enable end-to-end training with KKT-consistent gradients, offering a principled alternative to penalty tuning. Finally, operationalizing the framework with accelerated linear algebra (e.g., JAX) and releasing a reproducible codebase with nested cross-validation, ablations (no-RL, no-Bayes, no-HRP, sector-only), and formal inference would broaden applicability and facilitate independent verification.

\section{Conclusion}

This paper proposes a reinforcement–learning embedded Bayesian Hierarchical Risk Parity (RL-BHRP) framework that unifies three key desiderata of institutional portfolio construction: statistically stable inputs, interpretable risk budgeting, and adaptive allocation. Methodologically, we (i) impose a hierarchical Bayesian prior that shrinks asset expectations toward sector means and stabilizes covariance estimates; (ii) formalize a two–level weight map from sector budgets and within–sector allocations to final asset weights, with feasibility and interior bijection; (iii) derive a block–decomposed variance representation and an equivalence between sector–level risk parity and asset–level sector risk contributions; and (iv) embed this structure in a factorized softmax policy trained by policy gradients under an average–reward objective with a convex transaction–cost term and a hierarchical risk–parity dispersion penalty. Together, these elements yield an allocator that remains budget– and risk–aware while responding to time–varying opportunity sets.\\
\\
Empirically, the RL overlay improves both absolute and benchmark–relative performance over a 67–month out–of–sample window (Feb.\ 2020 to Aug.\ 2025). Relative to a static BHRP baseline and an equal–weight sector benchmark, RL--BHRP achieves the highest terminal wealth and geometric growth (CAGR \(15.16\%\) vs.\ \(13.37\%\) for BHRP and \(12.33\%\) for the benchmark), with the strongest risk–adjusted profile (Sharpe \(0.905\), Sortino \(1.648\)). Against the benchmark, RL--BHRP delivers an information ratio of \(0.687\) with low tracking error (\(3.69\%\)) and a positive Jensen alpha (\(2.76\%\)), indicating that the policy creates value while remaining benchmark–proximal. Drawdowns are comparable across all portfolios (RL trough \(-20.33\%\)), and the superior Calmar ratio reflects faster geometric recovery rather than reliance on excessive risk. Weight diagnostics reveal gradual, economically plausible tilts, with approximately half of the capital allocated to the top 15 names on average, high month–to–month persistence, and moderate turnover, consistent with the cost term in the reward.\\
\\
Substantively, these findings support the central thesis: hierarchical risk budgeting is a robust organizing principle for diversification, but coupling it with a Bayesian filter and a learned policy that selectively reallocates sector budgets and within–sector weights can harvest additional return without materially increasing active risk. The penalty on risk-contribution dispersion acts as an interpretable regularizer that connects the learned policy to a well–understood risk–parity target, thereby improving transparency and mitigating the usual black–box critique of RL in finance.\\
\\
The study has limitations. We work with adjusted prices from a public source and, for computational reasons, evaluate a sector–balanced subset of the mechanically screened universe; the data for the last month is partial. These choices do not compromise the integrity of the experimental design (no look–ahead, quarterly frozen membership, and tradability masks), but they motivate replication on survivorship-free, point–in–time databases and at a larger scale. Future research should (i) replace Gaussian likelihoods with heavy–tailed/dynamic specifications inside the Bayesian hierarchy; (ii) investigate exact risk–parity projections and alternative divergences for risk budgeting; (iii) adopt risk–sensitive RL objectives (distributional or CVaR–constrained actor–critic) under formal nonstationarity; and (iv) provide cross–fold inference (factor alphas, difference–in–Sharpe, SPA/MCS) alongside capacity and impact analysis.\\
\\
In conclusion, RL-BHRP offers a practical, theoretically grounded approach to adaptive, risk–aware portfolio construction. By combining hierarchical shrinkage with two–level risk parity and a constrained policy gradient, it achieves economically meaningful improvements in both absolute and relative terms, while preserving interpretability and control. The framework is modular and extensible, making it a promising baseline for future work on scalable, risk–sensitive learning in multi–asset portfolios.

\newpage
\bibliographystyle{plainnat} 
\bibliography{references} 

\newpage
\section*{Appendix A. Sector-Level Risk-Parity via Risk Budgeting: KKT System and Newton Solver}

Let \(G\ge 2\) be the number of sectors. The composite (sector) covariance is
\(\widetilde{\Sigma}(\eta)\in\mathbb{R}^{G\times G}\), symmetric positive definite, built from within‑sector weights \(\eta\) as in Eq.~\eqref{eq:SigmaTilde}.
We seek sector weights \(W\in\Delta_G=\{W\in\mathbb{R}^G_+:\mathbf{1}^\top W=1\}\) such that sector risk contributions are equal:
\[
\mathrm{RC}^{\mathrm{sec}}_g(W,\eta)=W_g\bigl(\widetilde{\Sigma}(\eta)W\bigr)_g
\quad\text{is constant in }g.
\]
Equivalently, \(\mathrm{RC}^{\mathrm{sec}}_g = b_g\,\sigma_p^2\) with budgets \(b_g>0\), \(\sum_g b_g=1\). For equal sector risk parity we take \(b_g=1/G\).

\subsection*{A.1 Risk-budgeting formulation and KKT conditions}

Consider the (scale‑invariant) risk‑budgeting program
\begin{equation}\label{eq:RB-program}
\min_{x\in\mathbb{R}^G_{++}}\;\; 
\frac{1}{2}\,x^\top \widetilde{\Sigma}(\eta)\, x
\quad\text{s.t.}\quad 
\sum_{g=1}^G b_g \log x_g \;=\; c,
\end{equation}
where \(c\in\mathbb{R}\) is an arbitrary constant fixing the geometric mean (any \(c\) yields the same solution up to scaling). Define the Lagrangian
\[
\mathcal{L}(x,\nu) \;=\; \tfrac{1}{2}x^\top \widetilde{\Sigma} x
\;-\; \nu\Bigl(\sum_{g} b_g \log x_g - c\Bigr).
\]
First‑order (KKT) conditions are
\begin{equation}\label{eq:KKT-RB}
\widetilde{\Sigma}x \;=\; \nu\, b \oslash x,
\qquad
\sum_g b_g \log x_g = c,
\qquad x\in\mathbb{R}^G_{++},
\end{equation}
where \(b=(b_g)_{g=1}^G\) and \(\oslash\) denotes componentwise division.

\subsubsection*{Risk‑budgeting \(\Rightarrow\) risk parity}\label{prop:RBtoRP}
Let \((x^\star,\nu^\star)\) satisfy the KKT system \eqref{eq:KKT-RB}. Then, with \(\sigma_p^2=x^{\star\top}\widetilde{\Sigma}x^\star\),
\[
\mathrm{RC}^{\mathrm{sec}}_g(x^\star,\eta)=x^\star_g\bigl(\widetilde{\Sigma}x^\star\bigr)_g
=\nu^\star b_g, 
\qquad
\sum_g \mathrm{RC}^{\mathrm{sec}}_g=\sigma_p^2=\nu^\star.
\]
Hence \(\mathrm{RC}^{\mathrm{sec}}_g=\;b_g\,\sigma_p^2\) for all \(g\).

Proof:
Multiply the \(g\)-th component of \(\widetilde{\Sigma}x^\star=\nu^\star b\oslash x^\star\) by \(x^\star_g\) to get 
\(\mathrm{RC}^{\mathrm{sec}}_g=\nu^\star b_g\). Summing over \(g\) and using \(\sum_g b_g=1\) yields \(\sum_g \mathrm{RC}^{\mathrm{sec}}_g=\nu^\star\).
But \(\sum_g \mathrm{RC}^{\mathrm{sec}}_g=x^{\star\top}\widetilde{\Sigma}x^\star=\sigma_p^2\) by conservation, hence \(\nu^\star=\sigma_p^2\) and the claim follows.

Remark: Normalization to the simplex
The solution \(x^\star\) of \eqref{eq:RB-program} is unique up to a positive scalar. The sector weights on the simplex are \(W^\star=x^\star/(\mathbf{1}^\top x^\star)\).
Risk budgets are homogeneous: for any \(s>0\), \(\mathrm{RC}^{\mathrm{sec}}_g(sx)=s^2\,\mathrm{RC}^{\mathrm{sec}}_g(x)\) and \(\sigma_p^2(sx)=s^2\,\sigma_p^2(x)\), so 
\(\mathrm{RC}^{\mathrm{sec}}_g/\sigma_p^2\) remains \(b_g\) after normalization.

\subsection*{A.2 Log-domain equations and Newton step}

Define \(y=\log x\in\mathbb{R}^G\) and \(u=\exp(y)\in\mathbb{R}^G_{++}\). Let \(z=\widetilde{\Sigma}u\).
The KKT stationarity \(\widetilde{\Sigma}u=\nu\,b\oslash u\) is equivalent to the componentwise equations
\begin{equation}\label{eq:log-eq}
f_i(y,c) \;=\; \log z_i + y_i - \log b_i - c \;=\; 0,\qquad i=1,\dots,G,
\end{equation}
where \(c=\log \nu\).
To fix the overall scale, impose the normalization constraint
\begin{equation}\label{eq:sum-constraint}
g(y) \;=\; \mathbf{1}^\top u - 1 \;=\; 0,
\end{equation}
so that the final \(x=\exp(y)\) already lies on the simplex.

\paragraph{Jacobian.}
Let \(U=\mathrm{diag}(u)\), \(Z=\mathrm{diag}(z)\).
The Jacobian of \(f(\cdot,c)\) with respect to \(y\) is
\begin{equation}\label{eq:J}
J(y) \;=\; I \;+\; Z^{-1}\,\widetilde{\Sigma}\,U,
\end{equation}
since \(\partial(\log z_i)/\partial y_j = (\widetilde{\Sigma}_{ij}u_j)/z_i\) and \(\partial y_i/\partial y_j = \delta_{ij}\).
The derivative of \(f\) with respect to \(c\) is \(-\mathbf{1}\).
The Jacobian of \(g\) with respect to \(y\) is \(\mathbf{1}^\top U\); with respect to \(c\) it is \(0\).

\paragraph{Augmented Newton system.}
At an iterate \((y^{(k)},c^{(k)})\), define the residuals 
\(r^{(k)} = f\bigl(y^{(k)},c^{(k)}\bigr)\) and \(s^{(k)} = g\bigl(y^{(k)}\bigr)\).
The Newton step \((\Delta y,\Delta c)\) solves the linear system
\begin{equation}\label{eq:newton-system}
\begin{bmatrix}
J(y^{(k)}) & -\mathbf{1}\\[2pt]
\mathbf{1}^\top U^{(k)} & 0
\end{bmatrix}
\begin{bmatrix}
\Delta y\\[2pt]
\Delta c
\end{bmatrix}
\;=\;
-\begin{bmatrix}
r^{(k)}\\[2pt]
s^{(k)}
\end{bmatrix},
\qquad
U^{(k)}=\mathrm{diag}\bigl(\exp(y^{(k)})\bigr).
\end{equation}
Update with backtracking to preserve positivity and decrease the residual:
\((y^{(k+1)},c^{(k+1)})=(y^{(k)},c^{(k)})+\alpha (\Delta y,\Delta c)\)
with \(\alpha\in(0,1]\) chosen by Armijo line search; positivity is automatic because \(x=\exp(y)\in\mathbb{R}^G_{++}\).

\subsubsection*{Existence, uniqueness, and global convergence}\label{thm:conv}
Assume \(\widetilde{\Sigma}(\eta)\succ 0\) and \(b\in\mathbb{R}^G_{++}\), \(\mathbf{1}^\top b=1\).
Then:
(i) There exists a unique \(x^\star\in\mathbb{R}^G_{++}\) (up to a scalar) satisfying the risk‑budgeting KKT conditions \eqref{eq:KKT-RB}.  
(ii) The log‑domain Newton method \eqref{eq:newton-system} with backtracking is globally convergent to \(y^\star=\log x^\star\) from any initial \(y^{(0)}\in\mathbb{R}^G\); the convergence is locally quadratic.

Proof
(i) The objective in \eqref{eq:RB-program} is strictly convex on \(\mathbb{R}^G_{++}\) and the constraint is affine in \(y=\log x\), hence the program has a unique minimizer for each \(c\), unique up to positive scaling in \(x\). The KKT system is therefore necessary and sufficient.  
(ii) In the log domain, \(f(\cdot,c)\) is smooth and its Jacobian \eqref{eq:J} is positive definite on the tangent space of the normalization constraint. Standard self‑concordance arguments (or Spinu’s log‑Newton analysis) yield global convergence with backtracking and quadratic local rate.

Special case (diagonal covariance)
If \(\widetilde{\Sigma}=\mathrm{diag}(\sigma_1^2,\dots,\sigma_G^2)\), then \((\widetilde{\Sigma}x)_g=\sigma_g^2 x_g\). The stationarity \(\sigma_g^2 x_g=\nu b_g/x_g\) gives \(x_g \propto \sqrt{b_g}/\sigma_g\). Normalization yields
\[
W_g^\star \;=\; \frac{\sqrt{b_g}/\sigma_g}{\sum_h \sqrt{b_h}/\sigma_h}.
\]
For equal budgets \(b_g=1/G\): \(W_g^\star \propto 1/\sigma_g\).

\subsection*{A.3 Practical algorithm}

\begin{algorithm}[H]
\caption{Sector risk‑parity via log‑domain Newton}
\begin{algorithmic}[1]
\State \textbf{Input:} \(\widetilde{\Sigma}\succ 0\), budgets \(b\in\mathbb{R}^G_{++}\), tolerance \(\epsilon\).
\State Initialize \(y^{(0)}=\mathbf{0}\), \(c^{(0)}=0\). \;(\(x^{(0)}=\mathbf{1}\), \(W^{(0)}=x^{(0)}/G\))
\Repeat
  \State \(u=\exp(y^{(k)})\), \(z=\widetilde{\Sigma}u\), \(U=\mathrm{diag}(u)\), \(Z=\mathrm{diag}(z)\).
  \State \(r=\log z + y^{(k)} - \log b - c^{(k)}\mathbf{1}\), \quad \(s=\mathbf{1}^\top u - 1\).
  \State Solve the linear system \eqref{eq:newton-system} for \((\Delta y,\Delta c)\).
  \State Backtracking line search on \(\alpha\in(0,1]\) to reduce \(\|r\|_2^2+s^2\); set \(y^{(k+1)}=y^{(k)}+\alpha\Delta y\), \(c^{(k+1)}=c^{(k)}+\alpha\Delta c\).
\Until{\(\|r\|_\infty \le \epsilon\) and \(|s|\le \epsilon\)} 
\State \textbf{Output:} \(W^\star=\exp(y^{(k)})\) (already on \(\Delta_G\)).
\end{algorithmic}
\end{algorithm}

\paragraph{Complexity and numerics.}
Each iteration solves a \((G{+}1)\times(G{+}1)\) linear system; cost is \(O(G^3)\), negligible for \(G\le 20\).
The method is scale‑free, strictly interior, and robust to conditioning when using double precision and mild regularization of \(\widetilde{\Sigma}\) (e.g., Ledoit–Wolf shrinkage at the asset level before aggregation).

\subsection*{A.4 Connection to the two-level HRP fixed point}

Given within‑sector weights \(\eta\), Theorem above shows
\(\sigma_p^2(W,\eta)=W^\top \widetilde{\Sigma}(\eta) W\) and
\(\mathrm{RC}^{\mathrm{sec}}_g(W,\eta)=W_g\bigl(\widetilde{\Sigma}(\eta)W\bigr)_g\).
Therefore the sector equal‑risk condition is precisely the risk‑budgeting KKT system \eqref{eq:KKT-RB} on \((W,\widetilde{\Sigma}(\eta))\) with budgets \(b_g=1/G\).
Inside each sector, an analogous log‑Newton or reciprocal fixed‑point update applies to \(\eta^{(g)}\) (cf.\ the within‑sector characterisation following Theorem~\ref{thm:decomp}).

\section*{Appendix B. Existence and Policy-Gradient Identity}

Let $(\mathcal{S},\mathcal{B}(\mathcal{S}))$ be a compact metric state space with Borel $\sigma$-algebra, and for each $s\in\mathcal{S}$ let $\mathcal{A}(s)\subset\mathbb{R}^m$ be a nonempty compact action set. The one-step reward $U:\mathcal{S}\times\mathbb{R}^m\to\mathbb{R}$ is bounded and continuous, and the transition kernel $P(\cdot\mid s,a)$ is \emph{Feller}: for every bounded continuous $f:\mathcal{S}\to\mathbb{R}$, the map $(s,a)\mapsto \int f(s')\,P(\mathrm{d}s'\mid s,a)$ is continuous. A (stationary, possibly randomized) policy $\pi$ is a stochastic kernel $\pi(\cdot\mid s)$ on $\mathcal{A}(s)$; for a parameterized policy we write $\pi_\theta(\cdot\mid s)$.

For any stationary policy $\pi$, define the \emph{induced kernel}
\[
P_\pi(\mathrm{d}s'\mid s) \;=\; \int_{\mathcal{A}(s)} P(\mathrm{d}s'\mid s,a)\,\pi(\mathrm{d}a\mid s),
\quad
\bar U_\pi(s) \;=\; \int_{\mathcal{A}(s)} U(s,a)\,\pi(\mathrm{d}a\mid s).
\]

\subsubsection*{Unichain, ergodicity, and invariance}\label{as:unichain}
For every stationary policy $\pi$, the Markov chain on $\mathcal{S}$ with kernel $P_\pi$ is $\psi$-irreducible, aperiodic, and positive recurrent, hence admits a unique invariant probability measure $d^\pi$ on $\mathcal{S}$; moreover, the chain is geometrically ergodic. 

\noindent
For such $\pi$, the long-run average reward exists and equals the stationary expectation
\[
\rho(\pi) \;:=\; \int_{\mathcal{S}} \bar U_\pi(s)\, d^\pi(\mathrm{d}s)
\;=\;\lim_{T\to\infty} \frac{1}{T}\,\mathbb{E}_\pi\!\Bigl[\sum_{t=0}^{T-1} U(s_t,a_t)\Bigr].
\]
For a parameterized policy we write $\rho(\theta):=\rho(\pi_\theta)$ and $J(\theta):=\rho(\theta)$.

\subsection*{B.1 Existence of an optimal stationary policy (average-reward)}

We establish existence via the vanishing-discount approach.

\subsubsection*{Lemma: Discounted problem - existence and measurability}\label{lem:disc}
For each $\beta\in(0,1)$, the discounted value function
\[
V_\beta(s) \;=\; \sup_\pi \; \mathbb{E}_\pi\!\Bigl[\sum_{t=0}^{\infty} \beta^t U(s_t,a_t)\,\Big|\,s_0=s\Bigr]
\]
solves the Bellman equation
\[
V_\beta(s) \;=\; \max_{a\in\mathcal{A}(s)} \Bigl\{ U(s,a) + \beta \int_{\mathcal{S}} V_\beta(s')\,P(\mathrm{d}s'\mid s,a)\Bigr\},
\]
and there exists a stationary \emph{deterministic} selector $\pi_\beta^\star$ attaining the maximum for all $s$. Moreover $V_\beta$ is bounded and continuous.

Proof:
Compactness of $\mathcal{A}(s)$, bounded continuity of $U$, and Feller continuity of $P$ imply the Bellman operator maps $C(\mathcal{S})$ into itself and is a contraction in the sup-norm. Berge's maximum theorem yields an upper hemicontinuous, nonempty argmax correspondence with measurable selectors; since the maximand is continuous and the correspondence compact-valued, a measurable \emph{deterministic} selector exists. Standard contraction arguments yield existence and uniqueness of $V_\beta\in C(\mathcal{S})$.

Define the \emph{vanishing-discount normalizations}
\[
\rho_\beta \;:=\; (1-\beta)\,V_\beta(s_0),\qquad
h_\beta(s) \;:=\; V_\beta(s) - V_\beta(s_0),
\]
for a fixed anchor $s_0\in\mathcal{S}$.

\subsubsection*{Limit points solve the ACOE}\label{lem:acoe}
There exists a sequence $\beta_n\uparrow 1$ such that $\rho_{\beta_n}\to \rho^\star\in\mathbb{R}$ and $h_{\beta_n}\to h^\star$ uniformly on $\mathcal{S}$, where $(\rho^\star,h^\star)$ solves the average-cost optimality equation (ACOE)
\[
\rho^\star + h^\star(s) \;=\; \max_{a\in\mathcal{A}(s)} \Bigl\{ U(s,a) + \int h^\star(s')\,P(\mathrm{d}s'\mid s,a)\Bigr\},\qquad s\in\mathcal{S}.
\]

Proof
By boundedness of $U$, the family $\{V_\beta\}_{\beta\in(0,1)}$ is equicontinuous and uniformly bounded; Ascoli–Arzelà yields relative compactness in $C(\mathcal{S})$. Hence there exists $\beta_n\uparrow 1$ with $h_{\beta_n}\to h^\star$ uniformly and $\rho_{\beta_n}\to \rho^\star$. Divide the discounted Bellman equation by $(1-\beta_n)$, subtract its value at $s_0$, and pass to the limit using dominated convergence and Feller continuity to obtain the ACOE.

\subsubsection*{Propsition: Existence of an average-reward optimal stationary policy}\label{prop:exist}
There exists a stationary deterministic policy $\pi^\star$ attaining $\rho^\star$, i.e.,
\[
\rho(\pi^\star) \;=\; \sup_\pi \rho(\pi) \;=\; \rho^\star,
\]
and $\pi^\star$ satisfies the ACOE: for each $s$, it selects an action in $\arg\max_{a\in\mathcal{A}(s)}\{ U(s,a)+\int h^\star(s')P(\mathrm{d}s'\mid s,a)\}$.

Proof:
By Lemma~\ref{lem:acoe} and Berge's maximum theorem, the ACOE maximizer admits a measurable selector $\pi^\star$; compactness of $\mathcal{A}(s)$ ensures the maximum is attained. Standard average-cost MDP theory (vanishing-discount limit) shows that any stationary selector that attains the ACOE is average-reward optimal. The unichain Assumption~\ref{as:unichain} ensures $\rho(\pi)$ is well-defined and independent of the initial state.

\subsection*{B.2 Policy-gradient identity (average reward)}

Fix a parameter $\theta$ and write $\pi=\pi_\theta$, $P_\theta$, $d^\theta$, $\rho(\theta)$. Define the \emph{differential value (bias) function} $h^\theta:\mathcal{S}\to\mathbb{R}$ as the (bounded) solution of the Poisson equation
\begin{equation}\label{eq:poisson}
h^\theta(s) \;=\; \sum_{t=0}^\infty \Bigl(\mathbb{E}_\theta[\,\bar U_\theta(s_t)\mid s_0=s\,] \;-\; \rho(\theta)\Bigr),
\qquad \int h^\theta(s)\, d^\theta(\mathrm{d}s)=0.
\end{equation}
Assumption~\ref{as:unichain} (geometric ergodicity on compact $\mathcal{S}$) and bounded $\bar U_\theta$ guarantee existence of a bounded solution $h^\theta$.

Define the $Q$-function
\begin{equation}\label{eq:Qdef}
Q^\theta(s,a) \;=\; U(s,a) \;-\; \rho(\theta) \;+\; \int h^\theta(s')\,P(\mathrm{d}s'\mid s,a).
\end{equation}
Note that $Q^\theta$ is bounded and measurable.

\subsubsection*{Theorem: Policy-gradient identity and finiteness}\label{thm:PG}
Suppose $U$ and $P$ do not depend on $\theta$, and $\pi_\theta(\cdot\mid s)$ is differentiable for $d^\theta$-almost every $s$, with $\int \|\nabla_\theta\log \pi_\theta(a\mid s)\|\,\pi_\theta(\mathrm{d}a\mid s)$ bounded uniformly in $s$. Then
\[
\nabla_\theta \rho(\theta) \;=\; \int_{\mathcal{S}} d^\theta(\mathrm{d}s)\; \int_{\mathcal{A}(s)} \pi_\theta(\mathrm{d}a\mid s)\;
\Bigl[\nabla_\theta \log \pi_\theta(a\mid s)\Bigr]\;Q^\theta(s,a),
\]
and the integral is finite.

Proof:
First write $\rho(\theta)=\int \bar U_\theta(s)\, d^\theta(\mathrm{d}s)$, where
$\bar U_\theta(s)=\int U(s,a)\,\pi_\theta(\mathrm{d}a\mid s)$.
Differentiate:
\begin{equation}\label{eq:drho1}
\nabla_\theta \rho(\theta) \;=\; \underbrace{\int (\nabla_\theta d^\theta)(\mathrm{d}s)\;\bar U_\theta(s)}_{T_1}
\;+\; \underbrace{\int d^\theta(\mathrm{d}s)\;\int U(s,a)\,\nabla_\theta \pi_\theta(\mathrm{d}a\mid s)}_{T_2}.
\end{equation}
We next eliminate $T_1$ via the stationary balance equation and the Poisson equation.

\emph{Step 1 (derivative of the invariant measure).}
The invariant measure satisfies $d^\theta(\cdot)=\int d^\theta(\mathrm{d}s)\,P_\theta(\cdot\mid s)$ with
$P_\theta(\cdot\mid s)=\int P(\cdot\mid s,a)\,\pi_\theta(\mathrm{d}a\mid s)$.
Differentiating the balance equation in the sense of measures gives
\[
(\nabla_\theta d^\theta)^\top (I - P_\theta) \;=\; (d^\theta)^\top \nabla_\theta P_\theta.
\]
Multiply on the right by the bounded solution $h^\theta$ to the Poisson equation
$(I - P_\theta) h^\theta = \bar U_\theta - \rho(\theta)\mathbf{1}$, and use $(\nabla_\theta d^\theta)^\top \mathbf{1}=\nabla_\theta 1=0$ to obtain
\begin{equation}\label{eq:keyid}
\int (\nabla_\theta d^\theta)(\mathrm{d}s)\;\bar U_\theta(s) \;=\; \int d^\theta(\mathrm{d}s)\,\int \nabla_\theta P_\theta(\mathrm{d}s'\mid s)\; h^\theta(s').
\end{equation}

\emph{Step 2 (expand the kernels).}
Because $U$ and $P$ do not depend on $\theta$,
\[
\nabla_\theta \bar U_\theta(s) \;=\; \int U(s,a)\,\nabla_\theta \pi_\theta(\mathrm{d}a\mid s),
\qquad
\nabla_\theta P_\theta(\mathrm{d}s'\mid s) \;=\; \int P(\mathrm{d}s'\mid s,a)\,\nabla_\theta \pi_\theta(\mathrm{d}a\mid s).
\]
Substitute these into \eqref{eq:drho1} and \eqref{eq:keyid}, and combine the two terms:
\[
\nabla_\theta \rho(\theta)
= \int d^\theta(\mathrm{d}s)\,\int \nabla_\theta \pi_\theta(\mathrm{d}a\mid s)\;
\Bigl[ U(s,a) + \int h^\theta(s')\,P(\mathrm{d}s'\mid s,a)\Bigr].
\]
Since $\int \nabla_\theta \pi_\theta(\mathrm{d}a\mid s) = \nabla_\theta 1 = 0$, we may subtract $h^\theta(s)$ inside the bracket without changing the integral. Using the definition \eqref{eq:Qdef} of $Q^\theta$ and the likelihood-ratio identity $\nabla_\theta \pi_\theta(\mathrm{d}a\mid s)=\pi_\theta(\mathrm{d}a\mid s)\,\nabla_\theta\log\pi_\theta(a\mid s)$,
\[
\nabla_\theta \rho(\theta)
= \int d^\theta(\mathrm{d}s)\,\int \pi_\theta(\mathrm{d}a\mid s)\,
\Bigl[\nabla_\theta \log \pi_\theta(a\mid s)\Bigr]\; Q^\theta(s,a).
\]
\emph{Finiteness.} Boundedness of $U$ and geometric ergodicity (Assumption~\ref{as:unichain}) imply the series defining $h^\theta$ converges absolutely and $h^\theta$ is bounded; hence $Q^\theta$ is bounded. The integrability condition on $\nabla_\theta\log \pi_\theta$ gives finiteness of the integral.

\paragraph{Conclusion.}
Proposition~\ref{prop:exist} establishes the existence of an optimal stationary policy for the average-reward criterion. Theorem~\ref{thm:PG} yields the policy-gradient identity
\[
\nabla_\theta J(\theta) \;=\; \mathbb{E}_{s\sim d^\theta,\,a\sim \pi_\theta(\cdot\mid s)}\!\bigl[\,\nabla_\theta \log \pi_\theta(a\mid s)\; Q^{\pi_\theta}(s,a)\,\bigr],
\]
with $Q^{\pi_\theta}$ defined by \eqref{eq:Qdef}. In our implementation, the factorized softmax policy satisfies the differentiability and integrability conditions, and the boundedness/Lipschitz properties of the reward established earlier guarantee the assumptions hold.

\end{document}